\documentclass[aps, prl, superscriptaddress, twocolumn, amsfonts, amsmath, amssymb, floatfix]{revtex4-1}

\usepackage{graphicx}
\usepackage{dcolumn}
\usepackage{bm}
\usepackage{epsfig}
\usepackage{braket}
\usepackage{color}
\usepackage{ulem}
\usepackage{amsmath}
\usepackage[colorlinks=true, letterpaper=true, pdfstartview=FitV, linkcolor=blue, citecolor=blue, urlcolor=blue]{hyperref}

\begin{document}

\title{Angular Momentum of a Bose-Einstein Condensate in a Synthetic Rotational Field}

\author{Chunlei Qu}
\email{chunleiqu@gmail.com}
\affiliation{INO-CNR BEC Center and Dipartimento di Fisica, Universit\`a di Trento, 38123 Povo, Italy}
\affiliation{JILA and Department of Physics, University of Colorado, Boulder, Colorado 80309, USA}

\author{Sandro Stringari}
\affiliation{INO-CNR BEC Center and Dipartimento di Fisica, Universit\`a di Trento, 38123 Povo, Italy}

\date{\today}

\begin{abstract}
By applying a position-dependent detuning to a spin-orbit-coupled Hamiltonian with equal Rashba and Dresselhaus coupling, we exploit the behavior of the angular momentum of a harmonically trapped Bose-Einstein condensed atomic gas and discuss the distinctive role of its canonical and spin components. By developing the formalism of spinor hydrodynamics, we predict the precession of the dipole oscillation caused by the synthetic rotational field, in analogy with the precession of the Foucault pendulum, the excitation of the scissors mode, following the sudden switching off of the detuning, and the occurrence of Hall-like effects. When the detuning exceeds a critical value we observe a transition from a vortex free, rigidly rotating quantum gas to a gas containing vortices with negative circulation which results in a significant reduction of the total angular momentum.  
\end{abstract}
	
\maketitle

\textcolor{blue}{\textit{Introduction.---}}
In recent years, the realization of artificial gauge fields has provided new opportunities for experimental and theoretical research in the field of cold atom physics~\cite{Goldman2014}. Synthetic gauge fields can be used to generate effective rotations, avoiding the difficulties associated with the rotation of the confining trap, as well as effective Lorentz forces acting on neutral atoms. A pioneering advance in the field was the experimental realization of quantized vortices~\cite{Lin2009}, by employing a pair of counter-propagating polarized laser beams which stimulate a Raman coupling between two different atomic hyperfine states and give rise to spin-orbit coupling with equal Rashba and Dresselhaus strengths~\cite{Lin2011}. The occurrence of rigidlike velocity patterns, corresponding to diffused vorticity and violating the irrotationality constraint of the superfluid velocity field~\cite{BECbook}, is another nontrivial feature predicted to occur in spin-orbit-coupled Bose-Einstein condensate (BEC)~\cite{Stringari2017}. Nowadays, the search for novel quantum effects caused by spin-orbit coupling in  many-body interacting systems is a subject of intensive investigations in the cold atom as well as in the condensed matter community.

Angular momentum is a quantity of fundamental importance in quantum many-body systems being directly related to their superfluid properties. In the presence of isotropic trapping, it, in fact, vanishes in regular superfluids at T=0, unless vortices are created at large enough angular velocities. In this Letter, we study the behavior of the angular momentum of a spin-orbit-coupled atomic gas in the presence of a position dependent detuning and explore its effects on the equilibrium and on the dynamics of the system. We explicitly reveal that spin-orbit-coupling strongly modifies the superfluid properties of the BEC and that angular momentum exhibits a nonzero value even before vortices are developed. Important consequences are the precession of the dipole oscillation, the excitation of the scissors mode, following the sudden switching off of the detuning as well as the emergence of Hall-like effects. For larger values of the detuning, a numerical simulation, based on the imaginary time evolution of the coupled Gross-Pitaevskii (GP) equations, reveals a transition from a vortex-free, rigidly rotating atomic gas to a phase consisting of many vortices with negative circulation \cite{Radic2011}.

We will consider the following single-particle Hamiltonian with an equal-Rashba-Dresselhaus spin-orbit coupling (for simplicity we set $\hbar=1$):
\begin{equation}
\hat{H}_{sp}=\frac{1}{2m}(\hat{p}_x-k_0\sigma_z)^2+\frac{\hat{p}_y^2}{2m} + V_{trap} -\frac{\Omega}{2}\sigma_x
-\eta k_0y\sigma_z
\label{eq:sp}
\end{equation}
where $k_0$ is the recoil momentum which is determined by the configuration of the Raman lasers, $V_{trap}=m(\omega_x^2x^2+\omega_y^2y^2)/2$ is the external trapping potential with $\omega_x$ and $\omega_y$ the oscillator frequencies along $x$ and $y$ directions. The dynamics along the $z$ direction is completely decoupled and, thus, is ignored hereafter. $\Omega>0$ is the Raman coupling strength, and $\eta$ is the coefficient of the position-dependent detuning term. Interaction effects will be taken into account through the mean field interaction term $V_\text{int}=(1/2)\sum_{\alpha\beta}\int d\mathbf{r}g_{\alpha\beta}n_\alpha n_\beta$ where $n_\alpha$ is the density distribution of the $\alpha$-th component, $g_{\alpha\beta}=4\pi a_{\alpha\beta}/m$ are the coupling constants in different spin channels and $a_{\alpha\beta}$ are the corresponding scattering lengths. The full Hamiltonian can be naturally employed in the framework of GP theory where the order parameter takes the form
\begin{equation}
\left(
\begin{array}{c}
\psi_1 \\
\psi_2
\end{array}
\right)=
\left(
\begin{array}{c}
\sqrt{n_1} e^{i\phi_1} \\
\sqrt{n_2} e^{i\phi_2}
\end{array}
\right).
\end{equation}
The mean-field approximation for the spin-orbit-coupled BEC is justified as quantum fluctuations are small for the realistic experimental parameters~\cite{Zheng2013}.

The angular momentum associated with the spin-orbit-coupled Hamiltonian is given by
\begin{equation}
\hat{L}_z=x\hat{p}_y-y(\hat{p}_x- k_0\sigma_z)=\hat{L}_z^c+\hat{L}_z^s
\end{equation}
where $\hat{L}_z^c=x\hat{p}_y-y\hat{p}_x$ is the canonical contribution, with $\hat{\mathbf{p}}=-i\hbar \mathbf{\nabla}$, while $\hat{L}_z^s=k_0y\sigma_z$ is the spin-dependent term. As discussed in Ref.~\cite{Stringari2017}, the detuning gradient $\eta$ introduced in Eq.~(\ref{eq:sp}) plays the role of an effective rotational frequency associated with the spin component $\hat{L}_z^s$ of the angular momentum.

\begin{figure}[t!]
\centerline{
\includegraphics[width=8.8cm]{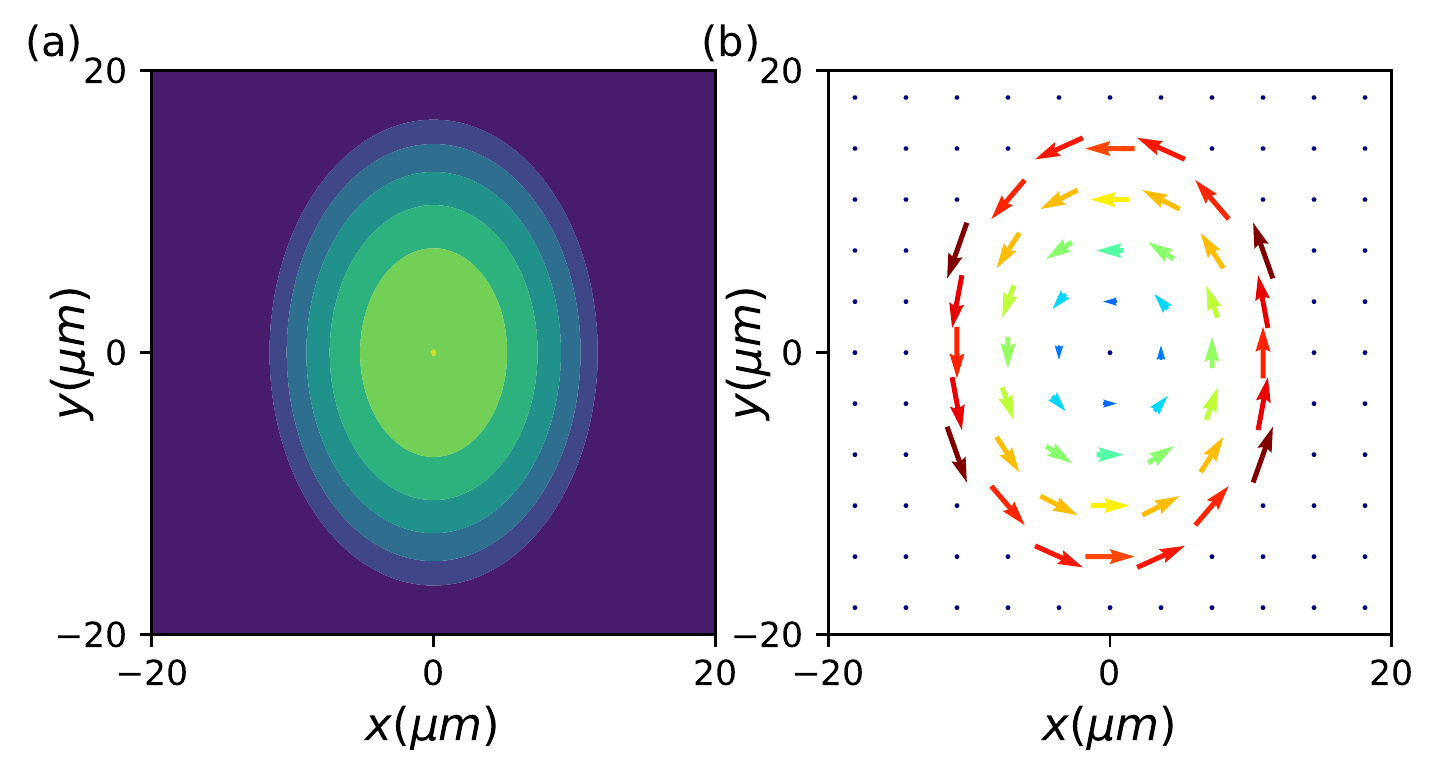}}
\caption{(a) Total density and (b) velocity field of a spin-orbit-coupled BEC confined in an anisotropic harmonic trap obtained from GP calculations, corresponding to $\Omega=8E_r$ ($E_r=k_0^2/2m=h\times 1775$Hz), $\eta=0.0025E_r$ and  trapping frequencies $(\omega_x,\omega_y)=2\pi\times(50\sqrt{2},50)$Hz, scattering length $a=100a_0$ where $a_0$ is the Bohr radius and the Thomas-Fermi radius $R_x^{TF}\approx 13.8\mu m$, $R_y^{TF}\approx 19.6\mu m$. For (a), the relative density is represented by colors with blue for zero density and yellow for maximum density. The size and color of the arrows in (b) reflects the magnitude of the velocity field. Red and long arrows correspond to large velocities while blue and short arrows correspond to small velocities.}
\label{Fig:anisotropic}
\end{figure}

\textcolor{blue}{\textit{Hydrodynamic theory.---}} For small values of the detuning parameter $\eta$, the coupled GP equations for the order parameters $\psi_{1,2}$ can be conveniently approximated by the following hydrodynamic form~\cite{Stringari2017,Qu2017}:
\begin{eqnarray}
&& \frac{\partial n}{\partial t} + \frac{1}{m}\nabla \cdot (n\nabla\phi) -\frac{k_0}{m}\nabla_x s_z
 = 0 \label{eq:n} \\
&& \frac{\partial\phi}{\partial t} + \frac{1}{2m}(\nabla\phi)^2
-\frac{\Omega}{2}\frac{n}{\sqrt{n^2-s_z^2}}+ng+V_{trap} = 0 \label{eq:phi} \\
&& 
-\frac{2k_0}{m}\nabla_x\phi +\Omega\frac{s_z}{\sqrt{n^2-s_z^2}} -2\eta k_0 y = 0 \label{eq:sz0}
\end{eqnarray}
where $n=n_1+n_2$ is the total density, $s_z=n_1-n_2$ is the spin density, and $\phi $ is the phase of the spinor order parameter. Equation (\ref{eq:n}) explicitly reveals the crucial role played by spin-orbit coupling in the equation of continuity, where the $x$ component of the current density $j_x=(n\nabla_x \phi -k_0s_z)/m$ contains a novel spin contribution. In deriving the spinor hydrodynamic equations, we have neglected the quantum pressure contribution to the kinetic energy terms, assumed $g_{\alpha\beta} \equiv g$ and taken into account the fact that, in the study of the collective modes oscillating with frequency satisfying  the condition $\omega_{coll} \ll \Omega$, the Raman coupling term is responsible for the locking of the relative phase of the two components, yielding $\phi_1=\phi_2 \equiv \phi$~\cite{Martone2012}. For simplicity, we have focused the discussion on the single-minimum phase, characterized by the condition $\Omega \ge \Omega_{c}$ where $\Omega_c=2k_0^2/m $ is the critical value of the Raman coupling at the transition from the single-minimum phase to the plane-wave phase \cite{Li2012a}. The ground state of the single-minimum phase is spin balanced while the low frequency excitations exhibit small spin density fluctuations ($s_z\ll n$); thus, Eqs.~(\ref{eq:phi}-\ref{eq:sz0}) can be simplified by the approximation $\sqrt{n^2-s_z^2}\approx n$. Consequently, the linearized spinor hydrodynamic equations can be written in the useful form
\begin{eqnarray}
&&\frac{\partial n}{\partial t} + \frac{1}{m^*}\nabla_x(n\nabla_x\phi)+\frac{1}{m}\nabla_y(n\nabla_y\phi)-\eta y\frac{\Omega_c}{\Omega}\nabla_x n =0  \nonumber \\ 
&& \label{eq:con} \\
&&\frac{\partial\phi}{\partial t} + \frac{1}{2m^*}(\nabla_x\phi)^2 + \frac{1}{2m}(\nabla_y\phi)^2
-\eta y\frac{\Omega_c}{\Omega} \nabla_x \phi-\frac{\Omega}{2}
 \nonumber \\
&& \qquad \qquad \qquad \qquad \qquad \qquad \qquad +ng+V_{trap} = 0
\end{eqnarray}
where $m^*=m(1-\Omega_c/\Omega)^{-1}$ defines the effective mass for the single minimum phase.

Adopting the Thomas-Fermi approximation for the total density $n=(\mu-V_{trap})/g$, where $\mu$ is the chemical potential, the  equilibrium solutions of the spinor hydrodynamic equations can be easily obtained by assuming the ansatz $\phi=\alpha xy$ and $s_z=2\beta yn$, yielding the results $\mathbf{v}^c=\alpha (y, x)/m$, $\mathbf{v}^s=2\beta k_0 (-y, 0)/m$ for the canonical and spin contributions to the velocity field ${\bf v}={\bf v}^c+{\bf v}^s={\bf j}/n$. One finds
\begin{equation}
\alpha=2\eta\frac{k_0^2}{\Omega}\frac{\omega_x^2}{\omega_{sc}^2}, \qquad \beta=\eta \frac{k_0}{\Omega}\frac{\omega_x^2+\omega_y^2}{\omega_{sc}^2},
\label{eq:alpha}
\end{equation}
where $\omega_{sc}=\sqrt{\omega_x^2(m/m^*)+\omega_y^2}$ is the scissors mode frequency in the absence of detuning ($\eta=0$).  The  canonical and spin components of the angular momentum take the simple form
\begin{eqnarray}
&\langle \hat{L}_z^c \rangle& = \langle x\nabla_y\phi-y\nabla_x\phi \rangle = \alpha\langle x^2-y^2\rangle 
\label{eq:CLz} \\
&\langle \hat{L}_z^s \rangle&  =\langle k_0ys_z/n \rangle  =2\beta k_0\langle y^2\rangle \; ,
\label{eq:Lz}
\end{eqnarray}
showing that, if the trapping frequencies satisfy the condition $\omega_x > \omega_y$, then $\langle x^2 \rangle <\langle y^2\rangle$, and thus, the canonical angular momentum (Eq.~\ref{eq:CLz}), originating from the irrotational component of the velocity field has the opposite sign to the spin angular momentum (Eq.~\ref{eq:Lz}) which is always positive if $\eta >0$. Here, we have used the notation $\langle \hat{A} \rangle=\int [An(x,y,z)]d\mathbf{r}$ to denote the average value of the operator $\hat{A}$. In the case of isotropic trapping the velocity field reduces to the rigid form $\mathbf{v}=\eta[\Omega_c/(2\Omega-\Omega_c)] \hat{e}_z\times \mathbf{r}$ and, since $\langle x^2 \rangle = \langle y^2 \rangle$, the canonical contribution to the  angular momentum identically vanishes~\cite{Stringari2017}. A typical density  and velocity field profile is presented in Fig.~\ref{Fig:anisotropic} where we have chosen the trapping frequencies $\omega_x=\sqrt{2}\omega_y$ and the Raman coupling $\Omega=2\Omega_c$. At equilibrium, the velocity field is always orthogonal to the gradient of the density, ensuring the stationarity of the profile.

\textcolor{blue}{\textit{Foucault precession.---}}In this section, we show that the effective rotational field associated with the position-dependent detuning, causes the precession of the dipole oscillation. In a regular BEC, the precession of the quadrupole oscillation, which has proven to be an efficient tool for revealing the presence of quantized vortices~\cite{Dalibard,Haljan2000} and for measuring the  moment of inertia of the system~\cite{Marago2000}, is directly related to the frequency splitting between the clockwise and counterclockwise quadrupole oscillations caused by the presence of angular momentum~\cite{Zambelli1998}. Such an effect is absent in the case of the dipole oscillation whose frequency, for Galilean invariant Hamiltonians, is independent of the angular momentum carried by the system and, consequently, does not exhibit precession in the laboratory frame. Since the position-dependent detuning introduces an effective rotational field, bringing the system into a noninertial frame, it is, as a result, interesting to explore its consequences on the precession of the dipole oscillation, in analogy with the Foucault precession exhibited by the classical pendulum in the noninertial frame of the rotating Earth~\cite{Landau}.

From the spinor hydrodynamic equations one can easily derive  coupled time-dependent differential equations for the average value of the dipole moment $\bar{x}_{\mu}(t)= \int (x_{\mu} n)d\mathbf{r}$ and of the canonical momentum ${\bar{p}_{\mu}}^{c}(t)=\int [(\nabla_{\mu} \phi) n] d\mathbf{r}$.  After eliminating the momentum variables, one finds the following equations:
\begin{eqnarray}
&&\frac{\partial^2 \bar{x}}{\partial t^2} + \left( (\omega_x^D)^2 - \frac{\eta^2}{4} \frac{m^*}{m}\frac{\Omega_c^2}{\Omega^2} \right)\bar{x} + \eta\frac{\Omega_c}{\Omega} \frac{\partial \bar{y}}{\partial t} = 0
\label{eq:Dx}  \\
&&\frac{\partial^2 \bar{y}}{\partial t^2} + \left( (\omega_y^D)^2 - \frac{\eta^2}{4}\frac{m^*}{m}\frac{\Omega_c^2}{\Omega^2} \right)\bar{y} - \eta \frac{m^*}{m}\frac{\Omega_c}{\Omega} \frac{\partial \bar{x}}{\partial t} = 0.
\label{eq:Dy}
\end{eqnarray}

If $\eta =0$, i.e., in the absence of detuning, the set of differential Eqs.~(\ref{eq:Dx}-\ref{eq:Dy}) predicted by spinor hydrodynamics admits the simple solutions $\omega^D_x=\omega_x\sqrt{m/m^*}$ and $\omega^D_y=\omega_y$ for the collective dipole frequencies along the $x$ and $y$ directions, respectively. The suppression of the dipole frequency $\omega_x^D$ with respect to the trap frequency $\omega_x$ is particularly large near the transition between the plane-wave phase and the single-minimum phase where $m^*\to \infty$ and $\omega_x^D\to 0$. This effect, predicted in Ref.~\cite{Li2012b}, was observed experimentally in Ref.~\cite{Zhang2012}. 

\begin{figure}[t!]
\centerline{
\includegraphics[width=8.6cm]{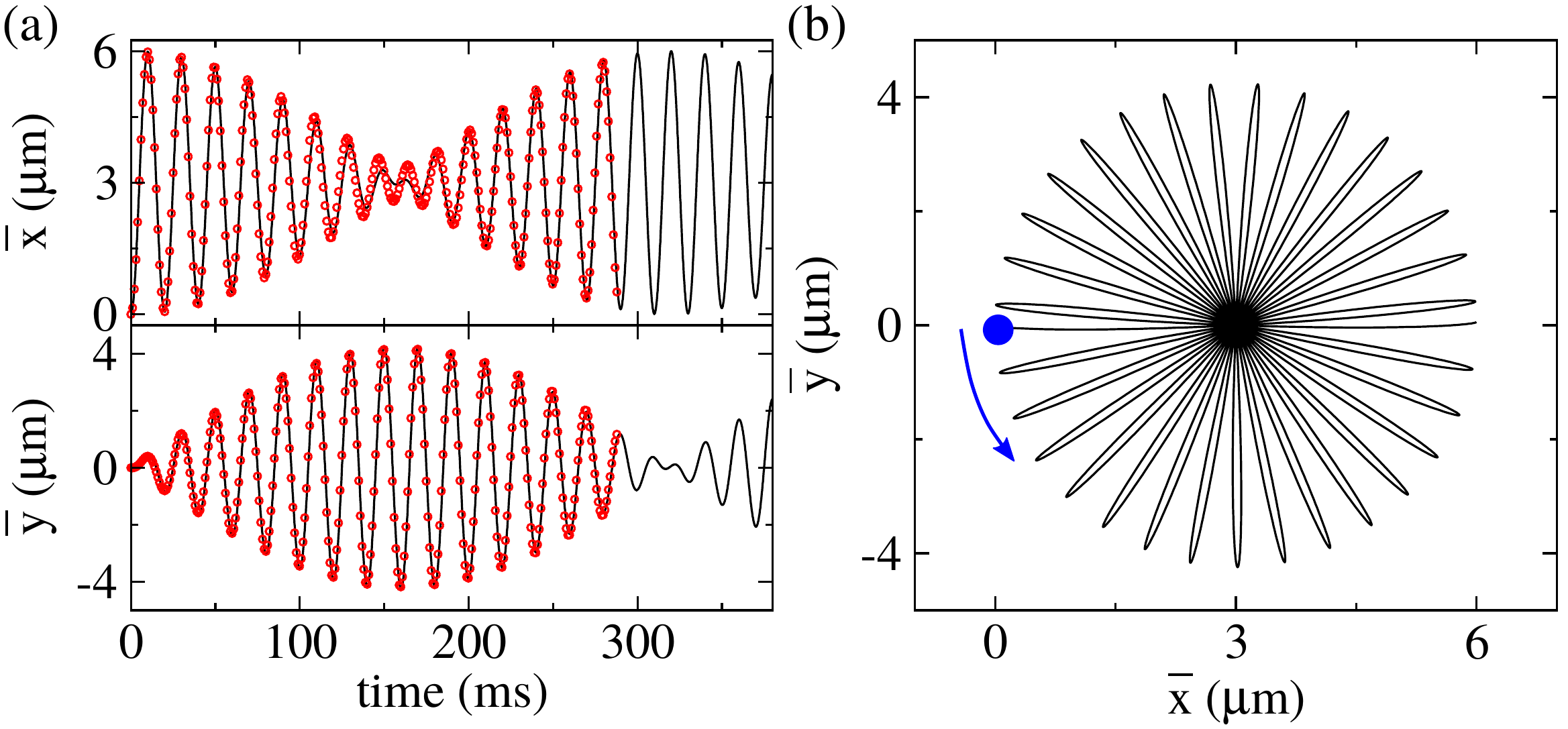}}
\caption{Foucault precession of the spin-orbit-coupled BEC. (a) Time-dependence of the center-of-mass position $(\bar{x}, \bar{y})$ of the spin-orbit-coupled Bose gas in the presence of a position dependent detuning. The dynamics is initiated by exciting a dipole mode along the $x$ direction, achieved by suddenly shifting the trap center to $x_0=3\mu m$.  The black lines are the hydrodynamic results and the red dots are the results obtained from the GP calculation. (b) Precession of the dipole oscillation. The starting point of the motion is $(0,0)$ (blue dot) and the blue arrow indicates the precession direction. The system parameters are the same as in Fig.~\ref{Fig:anisotropic}.}
\label{Fig:Foucault}
\end{figure}

If $\eta\neq 0$, the precession effect of the dipole oscillation is best revealed when the two unperturbed dipole frequencies are degenerate, i.e. when $\omega_x^D=\omega_y^D\equiv \omega_D$. This condition is easily achieved by properly choosing the values of the oscillator frequencies for a given value of the Raman coupling. Since $m^*/m$ is always larger than unity, it follows that the trapping frequency $\omega_x$ must be larger than $\omega_y$.  In the presence of degeneracy, the position-dependence detuning in the spin-orbit Hamiltonian causes an important coupling between the two unperturbed dipole modes, yielding a typical beating effect~\cite{note}. 

The resulting predictions are reported in Fig.~\ref{Fig:Foucault}. The coupling between the unperturbed dipole modes results into renormalized counterclockwise ($\omega_+$) and clockwise ($\omega_-$) solutions with frequencies given by 
\begin{equation}
\omega_{\pm}=\omega_D \pm \frac{\eta}{2}\sqrt{\frac{m^*}{m}}\frac{\Omega_c}{\Omega},
\end{equation}
corresponding to the oscillating solutions of the dipole moments $x\pm i \sqrt{m/m^*}y$. The splitting between the two frequencies coincides with the precession frequency  $\omega_{prec}=\omega_+-\omega_-= \eta\sqrt{\frac{m^*}{m}}\frac{\Omega_c}{\Omega}$ of the dipole oscillation in the $x-y$ plane. Notice that the maximum amplitudes of the oscillations along the $x$ and $y$ directions do not coincide due to the fact that, despite the two unperturbed frequencies being the same, the effective mass is larger along the $x$ direction. The ratio of the maximum oscillation amplitudes along the $y$ and $x$ directions is given by $\sqrt{m^*/m}$. We have verified that our analytic predictions, and, in particular, the value of the precession frequency  $\omega_{prec}$, coincide with high precision with the numerical solutions of the coupled GP equations (red dots in Fig.~\ref{Fig:Foucault}(a))~\cite{note1}. 

\begin{figure}[t!]
\centerline{
\includegraphics[width=8.6cm]{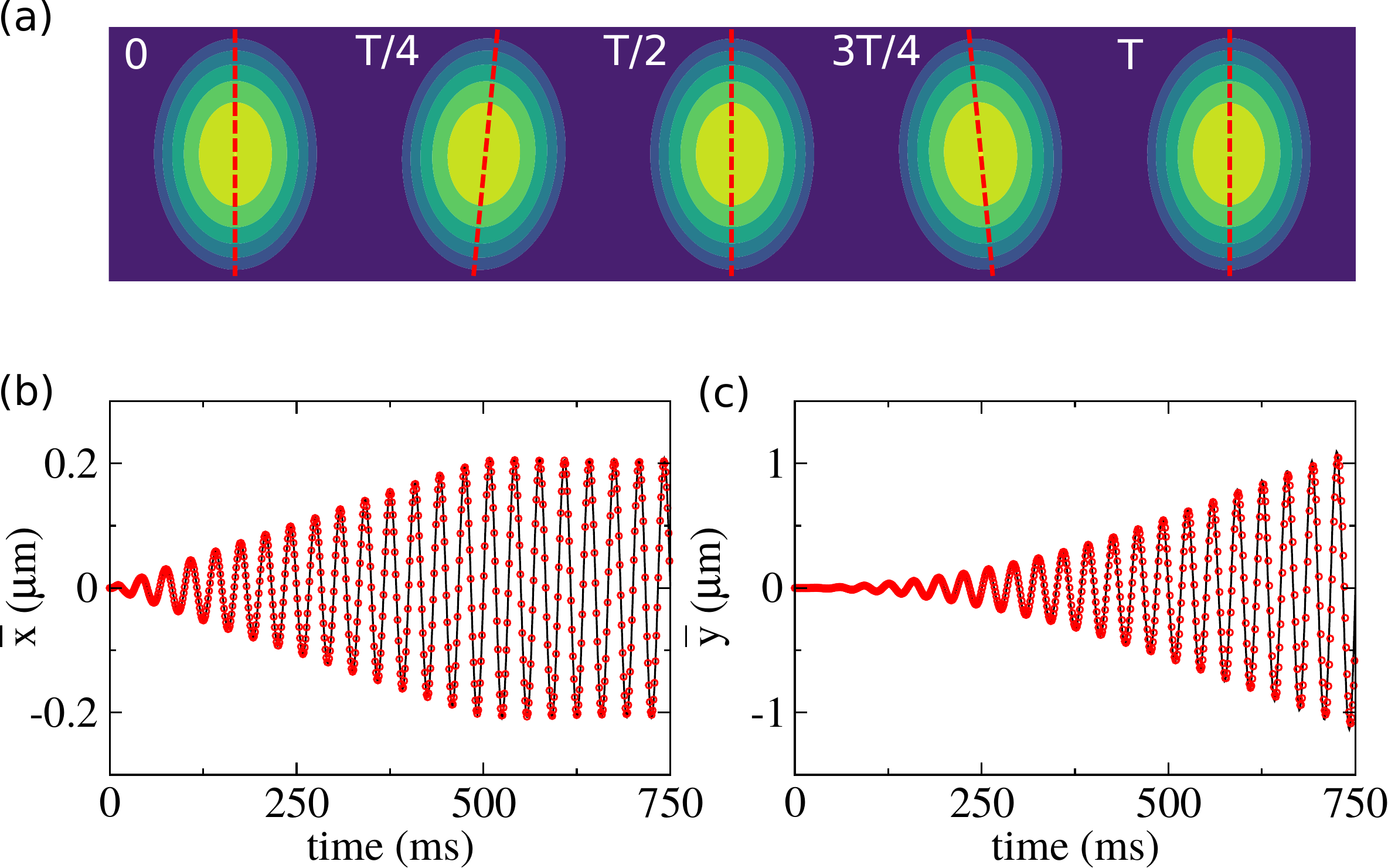}}
\caption{(a) Excitation of the scissors mode after turning off the position dependent detuning. The red dashed lines are the long symmetric axes of the BEC and $T=2\pi/\omega_{sc}$ is the period of the scissors mode. The other system parameters are the same as in Fig.~\ref{Fig:anisotropic}. The relative density is represented by colors with blue for zero density and yellow for maximum density. (b,c) Excitation of the Hall-like effect. A periodic modulation of the trap center  $x_0(t)=A(t)\sin(\omega t)$ along the $x$ direction causes a resonant response of the dipole oscillation along the $y$ direction if $\omega=\omega_y$. The amplitude $A(t)$ is, here, linearly ramped to the final value $A=0.18\mu m$ at $t=500ms$ and then hold for another $250ms$.  The dipole oscillations along $x$ and $y$ are shown in (b) and (c) respectively where the black solid lines are the hydrodynamic prediction and the red dots are the results from the GP simulation. The trap frequencies are $(\omega_x,\omega_y)=2\pi\times(120,30)Hz$, the Raman coupling $\Omega=8E_r$ and the detuning gradient $\eta=0.002E_r$.}
\label{Fig:Hall}
\end{figure}

\textcolor{blue}{\textit{Scissors mode and Hall effect.---}} Another easily measurable effect caused by the presence of angular momentum is the excitation of the scissors mode, following the sudden switching off of the position dependent detuning. In this case, the canonical angular momentum carried by the system (Eq.~\ref{eq:CLz}) causes the initial rotation of the cloud, which then oscillates at the scissors mode frequency $\omega_{sc}$ (see Fig.~\ref{Fig:Hall}(a)). The density perturbation of the cloud can be analytically obtained from our hydrodynamic theory: 

\begin{equation}
\delta n(t)=\eta\frac{m}{\hbar g}\frac{\Omega_c}{\Omega}\frac{\omega_x^2}{\omega_{sc}}\sin(\omega_{sc}t)xy,
\nonumber
\end{equation}
which is valid for both the anisotropic and the isotropic geometries.

When the two unperturbed dipole oscillation frequencies are different (i.e, $\omega_x^D\neq \omega_y^D$), the spin-orbit configuration is well suited to investigating a Hall-like effect~\cite{Hall1879} associated with the appearance of a current along the $y$ direction caused by a force applied to the $x$-direction. One can periodically modulate the external harmonic trapping potential along the $x$-direction with a frequency in resonance with the frequency $\omega_y$ of the dipole oscillation along the $y$-direction. After a few oscillations, the coupling induced by the detuning is responsible for a huge excitation of the dipole oscillation along the $y$ direction, while it practically does not affect the motion along $x$ (see Fig.~\ref{Fig:Hall}(b,c)). A Hall effect of similar type was observed experimentally in Ref.~\cite{LeBlanc2012} by exciting the axial compression mode with a frequency in resonance with the scissors mode.

\begin{figure}[t!]
\centerline{\includegraphics[width=8.0cm]{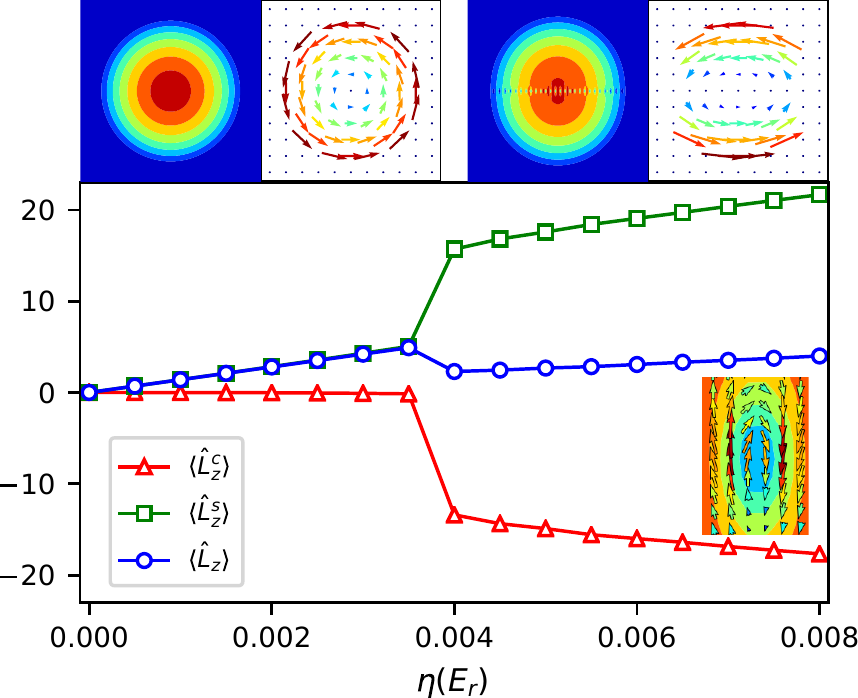}}
\caption{Dependence of the total angular momentum $\langle \hat{L}_z\rangle $ (blue circles), the canonical $\langle \hat{L}_z^c\rangle$ (red triangles), and the spin-dependent $\langle \hat{L}_z^s\rangle$ (green squares) contributions on the detuning gradient $\eta$ of a spin-orbit-coupled atomic gas in an isotropic harmonic trap with $(\omega_x,\omega_y)=2\pi\times(50,50)Hz$ and the Thomas-Fermi radius $R^{TF}\approx 17.9\mu m$. The results are obtained by numerical solving of the GP equations. The Raman coupling strength is $\Omega=\Omega_c$. Top panels show typical density profile and velocity field of the two phases blow and above $\eta_c$. The inset panel shows the velocity field overlapped with the density profile near the center of a vortex. For contour plots, the relative density is represented by colors with blue for zero density and red for maximum density.}
\label{Fig:Lz}
\end{figure}

\textcolor{blue}{\textit{Production of vortices.---}}
In this section we consider, for the sake of simplicity, only the case of isotropic trapping and the results correspond to the ground state of the system obtained by an imaginary time evolution of the coupled GP equations at $\Omega=\Omega_c$. Above a critical value, $\eta_c$ of the detuning, we observe a first-order phase transition from vortex-free atomic gas exhibiting a rigid rotation \cite{Stringari2017}, to a phase consisting of many vortices. The distinctive features exhibited by the two phases are caused by the fact that the angular momentum of the spin-orbit-coupled atomic gas includes both the canonical and the spin-dependent contribution. For small detuning ($\eta<\eta_c$), the canonical angular momentum vanishes~\cite{note2}, and the spin-dependent term gives rise to a finite angular momentum which increases linearly as a function of $\eta$, reflecting the rigid value of the moment of inertia. For large detuning ($\eta>\eta_c$), vortices are developed in the system and the corresponding canonical angular momentum becomes nonzero and has the opposite sign with respect to the spin dependent term, resulting in a significant reduction of the total angular momentum (see Fig.~\ref{Fig:Lz}). Above the critical value of the detuning, a line of vortices appears at $y=0$ (Fig.~\ref{Fig:Lz}, top panels) which separates the system into two parts corresponding to two dressed states with quasimomentum $\pm q_x$~\cite{note3}. This is directly revealed by the canonical velocity field. The superfluid flows in opposite directions above and below the line of vortices. As explicitly shown in the inset of Fig.~\ref{Fig:Lz}, these vortices have negative circulation, as first pointed out in \cite{Radic2011}.

In summary, we have shown that the angular momentum of a spin-orbit-coupled BEC exhibits a deeply different behavior with respect to the case of a regular BEC, which is characterized by the irrotational constraint for the velocity field. In addition to the appearance of crucial rigid rotational components, we have shown that the presence of a position dependent detuning brings the system into an effective rotating frame, causing the precession of the dipole oscillation, the possibility of exciting the scissors mode, and the observation of the Hall-like effect. We have explicitly discussed how the competition between the canonical and the spin components of the angular momentum affects the behavior of the system both at small and high values of the detuning when vortices of negative circulation are observed. We expect that our predictions will stimulate further experimental and theoretical work on the intriguing behavior of angular momentum in quantum many-body systems in the presence of artificial gauge fields.

\begin{acknowledgments}
We thank Ana M. Rey, Murray Holland, and Chuan-Hsun Li for carefully reading the manuscript. This project has received funding from the European Union's Horizon 2020 research and innovation programme under Grant Agreement No. 641122 ``QUIC" and the Instituto Nazionale di Fisica Nucleare.
\end{acknowledgments}


\begin{thebibliography}{52}%
\bibitem{Goldman2014} N. Goldman, G. Juzeliunas, P. Ohberg, and I. B. Spielman, \textit{Light-induced gauge fields for ultracold atoms}, \href{https://doi.org/10.1088/0034-4885/77/12/126401}{Rep. Prog. Phys. \textbf{77} 126401 (2014).}

\bibitem{Lin2009} Y.-J. Lin, R. L. Compton, K. Jim\'enez-Garc\'ia, J. V. Porto, and I. B. Spielman, \textit{Synthetic magnetic fields for ultracold neutral atoms}, \href{https://doi.org/10.1038/nature08609}{Nature \textbf{462}, 628 (2009).}

\bibitem{Lin2011} Y.-J. Lin, K. Jim\'enez-Garc\'ia, and I. B. Spielman, \textit{Spin-orbit-coupled Bose–Einstein condensates}, \href{https://doi.org/10.1038/nature09887}{Nature (London) \textbf{471}, 83 (2011).}

\bibitem{BECbook} L. Pitaevskii and S. Stringari, \textit{Bose-Einstein condensation and superfluidity} (Oxford University Press, Oxford, England, 2016).

\bibitem{Stringari2017} S. Stringari, \textit{Diffused Vorticity and Moment of Inertia of a Spin-Orbit Coupled Bose-Einstein Condensate}, \href{https://doi.org/10.1103/PhysRevLett.118.145302}{Phys. Rev. Lett. \textbf{118}, 145302 (2017).}

\bibitem{Radic2011} J. Radi\'c, T. A. Sedrakyan, I. B. Spielman, and V. Galitski, \textit{Vortices in spin-orbit-coupled Bose-Einstein condensates}, \href{https://doi.org/10.1103/PhysRevA.84.063604}{Phys. Rev. A \textbf{84}, 063604 (2011).}

\bibitem{Zheng2013} W. Zheng, Z.-Q. Yu, X. Cui, and H. Zhai, \textit{Properties of Bose gases with the Raman-induced spin-orbit coupling}, \href{https://doi.org/10.1088/0953-4075/46/13/134007}{J. Phys. B: At. Mol. Opt. Phys. \textbf{46}, 134007 (2013).}

\bibitem{Qu2017} C. Qu, L. P. Pitaevskii, and S. Stringari, \textit{Spin-orbit-coupling induced localization in the expansion of an interacting Bose Einstein condensate}, \href{https://doi.org/10.1088/1367-2630/aa7e8c}{New J. Phys. \textbf{19}, 085006 (2017).}

\bibitem{Martone2012} G. I. Martone, Y. Li, L. P. Pitaevskii, and S. Stringari, \textit{Anisotropic dynamics of a spin-orbit-coupled Bose-Einstein condensate}, \href{https://doi.org/10.1103/PhysRevA.86.063621}{Phys. Rev. A \textbf{86}, 063621 (2012).}

\bibitem{Li2012a} Y. Li, L. P. Pitaevskii, and S. Stringari, \textit{Quantum Tricriticality and Phase Transitions in Spin-Orbit Coupled Bose-Einstein Condensates}, \href{https://doi.org/10.1103/PhysRevLett.108.225301}{Phys. Rev. Lett. \textbf{108}, 225301 (2012)}

\bibitem{Dalibard} K. W. Madison, F. Chevy, W. Wohlleben, and J. Dalibard, \textit{Vortex Formation in a Stirred Bose-Einstein Condensate}, \href{https://doi.org/10.1103/PhysRevLett.84.806}{Phys. Rev. Lett. \textbf{84}, 806 (2000).}

\bibitem{Haljan2000} P. C. Haljan, B. P. Anderson, I. Coddington, and E. A. Cornell, \textit{Use of Surface-Wave Spectroscopy to Characterize Tilt Modes of a Vortex in a Bose-Einstein Condensate}, \href{https://doi.org/10.1103/PhysRevLett.86.2922}{Phys. Rev. Lett. \textbf{86}, 2922 (2001).}

\bibitem{Marago2000} O. M. Marag\'o, S. A. Hopkins, J. Arlt, E. Hodby, G. Hechenblaikner, and C. J. Foot, \textit{Observation of the Scissors Mode and Evidence for Superfluidity of a Trapped Bose-Einstein Condensed Gas}, \href{https://doi.org/10.1103/PhysRevLett.84.2056}{Phys. Rev. Lett. \textbf{84}, 2056 (2000).}

\bibitem{Zambelli1998} F. Zambelli and S. Stringari, \textit{Quantized Vortices and Collective Oscillations of a Trapped Bose-Einstein Condensate}, \href{https://doi.org/10.1103/PhysRevLett.81.1754}{Phys. Rev. Lett. \textbf{81}, 1754 (1998).}

\bibitem{Landau} L. D. Landau and E. M. Lifshitz, \textit{Mechanics}, 3rd ed. (Butterworth-Heinemann, London, 1976).

\bibitem{Li2012b} Y. Li, G. I. Martone, and S. Stringari, \textit{Sum rules, dipole oscillation and spin polarizability of a spin-orbit coupled quantum gas}, \href{https://doi.org/10.1209/0295-5075/99/56008}{Europhys. Lett. \textbf{99}, 56008 (2012).}

\bibitem{Zhang2012} J.-Y. Zhang, S.-C. Ji, Z. Chen, L. Zhang, Z.-D. Du, B. Yan, G.-S. Pan, B. Zhao, Y.-J. Deng, H. Zhai, S. Chen, and J.-W. Pan, \textit{Collective Dipole Oscillations of a Spin-Orbit Coupled Bose-Einstein Condensate}, \href{https://doi.org/10.1103/PhysRevLett.109.115301}{Phys. Rev. Lett. \textbf{109}, 115301 (2012).} 

\bibitem{note} A similar beating effect of the dipole modes has been previously studied in the absence of rotational effects,  as a consequence of the  spin-dependent interaction term in the Hamiltonian ~\cite{Chen2012}. 

\bibitem{Chen2012} Z. Chen and H. Zhai, \textit{Collective-mode dynamics in a spin-orbit-coupled Bose-Einstein condensate}, \href{https://doi.org/10.1103/PhysRevA.86.041604}{Phys. Rev. A \textbf{86}, 041604(R) (2012).}

\bibitem{note1} Only for large amplitude oscillations, deviations are visible because the linearized hydrodynamic theory adopts the value of the  effective mass calculated at zero quasimomentum and cannot consequently capture the large-amplitude dynamics.



\bibitem{Hall1879} E. H. Hall, \textit{On a New Action of the Magnet on Electric Currents}, \href{https://doi.org/10.2307/2369245}{Am. J. Math. \textbf{2}, 287 (1879).}

\bibitem{LeBlanc2012} L. J. LeBlanc, K. Jim\'enez-Garc\'ia, R. A. Williams, M. C. Beeler, A. R. Perry, W. D. Phillips, and I. B. Spielman, \textit{Observation of a superfluid Hall effect}, \href{www.pnas.org/cgi/doi/10.1073/pnas.1202579109}{Proc. Natl. Acad. Sci. U.S.A. \textbf{109}, 10811 (2012).}

\bibitem{note2} Though the trap is isotropic, the condensate is not strictly isotropic due to the presence of the $y$-position dependent detuning. The size of the cloud along $y$ is slightly larger than that along $x$, and thus, the canonical angular momentum is not exactly zero, but exhibits a pretty small negative value for $\eta < \eta_c$ (less than $3\%$ of the total angular momentum in Fig.~\ref{Fig:Lz}). We emphasize that, for general trap geometries, the canonical angular momentum does not vanish, but the appearance of vortices array at a large enough detuning gradient is similar.

\bibitem{note3} For $\Omega>\Omega_c$, vortices do not necessarily appear at $y=0$ and  distribute over all the condensate, the behavior of angular momentum being, however, similar.

\end{thebibliography}
\end{document}